# Millimeter-scale wide-field mid-infrared photothermal imaging enabled by a broadly tunable picosecond optical parametric oscillator

A. Thayyil Raveendran[1], S. Cael[1], A. Jose[2], J. Popp[1,3], and C. Krafft[1,†]

[1]Leibniz Institute of Photonic Technology, Member of Leibniz Research Alliance Leibniz Health Technologies, Member of the Leibniz Center for Photonics in Infection Research, Albert-Einstein-Str. 9, 07745 Jena, Germany

[2]Otto Schott Institute of Materials Research, Friedrich-Schiller-University Jena, Lessingstrasse 14, 07743 Jena, Germany

[3]Friedrich-Schiller-University Jena, Institute of Physical Chemistry, Member of Leibniz Research Alliance Leibniz Health Technologies, Member of the Leibniz Center for Photonics in Infection Research, Helmholtzweg 4, 07743 Jena, Germany
*Christoph.Krafft@leibniz-ipht.de

**Abstract:** Wide-field mid-infrared photothermal (MIP) imaging enables chemically specific microscopy with submicron spatial resolution but remains fundamentally limited by the trade-off between field of view, mid-infrared pulse energy, and spectral tunability. As a result, current wide-field implementations are typically restricted to fields of view below 200 µm and to either the fingerprint or high-wavenumber spectral regions. Here, we overcome these limitations by developing a wide-field fluorescence-detected mid-infrared photothermal (F-MIP) imaging platform driven by a commercial picosecond optical parametric oscillator (OPO). The system provides pulse energies of up to 360 µJ together with a broad tuning range from 625 to 4327 $cm^{-1}$, enabling millimeter-scale wide-field imaging in the high-wavenumber regions. We demonstrate a field of view of approximately 1 mm in diameter for fluorescently labeled polystyrene beads while preserving spectral fidelity. Furthermore, the platform enables, to our knowledge, the first wide-field MIP imaging below 900 $cm^{-1}$. To demonstrate its applicability to biomedical imaging, we performed large-area mosaic imaging of fluorescent tuberculosis-infected tissue sections, providing chemically resolved maps over millimeter-sized sample areas. These results establish broadly tunable OPO-driven F-MIP as a scalable platform for high-throughput vibrational imaging of large biological specimens and advanced materials.

Mid-infrared photothermal (MIP) imaging provides sensitive, label-free chemical contrast by exploiting vibrational absorption to induce detectable refractive-index and volume changes in a sample, enabling sub-micron spatial resolution[1]. MIP imaging has primarily relied on single-point or raster-scan excitation with photodiode detection, while recent implementations employ quantum cascade laser (QCL)-based laser scanning for high-speed imaging[2, 3]. However, such laser-scanning methods require precise optical alignment, and are constrained in high-speed imaging by the finite deflection range of scanning mirrors. In contrast, the incorporation of high-speed CMOS array detectors enables simultaneous measurement of the probe light across many pixels, thereby providing single-shot acquisition over larger areas on millisecond timescales[4]. Recently, wide-field MIP imaging has been realized through various optical detection schemes combined with QCL or optical parametric oscillators (OPO) sources[4, 5], yet the achievable field of view (FOV) and spectral tuning range remain constrained by the properties of available mid-infrared sources.

Early wide-field MIP implementations using pulsed OPO sources in the fingerprint region demonstrated label-free chemical imaging of microscale samples over FOVs of ~35 µm[6]. Quantitative phase imaging with QCL-driven MIP achieved a ~100 µm FOV through phase-sensitive readout at reduced mid-IR fluence[7], while heterodyne detection with frequency-domain lock-in readout attained comparable FOV extension through improved detection sensitivity[8]. Further FOV extension to ~180 µm was achieved using high-pulse-energy (µJ-range) OPO sources, where short-pulse, nanosecond excitation generates rapid localized heating and larger transient temperature rises compared to longer-pulse regimes, thereby enhancing the photothermal signal[1, 9, 10]. Moreover, FOV scalability is governed by mid-IR fluence: distributing pulse energy over a larger area reduces local fluence and degrades modulation depth, making high-pulse-energy sources essential for large-area imaging[3]. Fluorescence-detected MIP (F-MIP) yields a fluorescence modulation depth roughly 100 times larger than scattering-based detection [11, 12]. Subsequently, the combination of fluorescence-sensitivity detection and high-pulse-energy mid-IR sources has the potential to achieve chemical-contrast imaging with an expanded FOV.

In this study, we introduce wide-field F-MIP imaging as a platform technology for large-area, chemically specific imaging, enabled by a commercial OPO laser source. The OPO delivers pulse energies up to 360 µJ with pulse widths ~26 ps at a 100 Hz repetition rate and offers broad tunability across 625–4347 $cm^{-1}$. Using this mid-IR source, we performed wide-field MIP imaging of polystyrene (PS) beads and tuberculosis-infected tissue sections across both the fingerprint and high-wavenumber spectral regions, demonstrating the first wide-field MIP platform to combine millimeter-scale FOV with broad spectral tunability spanning the fingerprint region to high-wavenumber bands. This breakthrough enables large-area, high-speed mid-IR chemical imaging with applications spanning biological tissue analysis and advanced materials characterization.

The wide-field F-MIP system includes an upper optical path for probing (Fig. 1) and a lower path for mid-IR excitation. A 450 nm LED (UHPT-450-SR, Pritzmatrix) is used as the probe light source to excite fluorescence in the sample. The illumination is relayed to the back focal plane of the objective lens (10×/0.25 NA, 20×/0.4 NA, or 50×/0.75 NA) using a converging lens, providing uniform wide-field excitation. The excitation light is directed to the sample by a dichroic mirror (MD480, Thorlabs). Fluorescence emitted from the sample is collected by the same objective, transmitted through the dichroic mirror, filtered using an emission filter (FELH0500, Thorlabs), and imaged onto a CMOS camera (MX245MG-SY-X4G3-FF, Ximea) via a tube lens. The mid-IR excitation path employs reflective optics. The 100-Hz pulsed mid-IR beam (PT501, Ekspla) is expanded, reduced, or maintained at the same diameter depending on the mid-IR wavelength. The beam is then guided to the sample through a series of gold mirrors (PF10-03-M02, Thorlabs) and focused onto the sample plane using an additional off-axis parabolic mirror 1 (OAP1) (MPD029-P01, MPD169-P01, or MPD189-P01, Thorlabs). OAP2 and OAP3 (MPD139-M01 and MPD169-P01, Thorlabs) tailor the beam size, and OAP1 then tightly focuses the resulting beam onto the sample plane. The OAP1 and mirror M1 are mounted on a linear translation stage, allowing precise adjustment of the focal position. The optical configuration was adjusted in relation to both the required FOV and spectral range. OAP2 and OAP3 were paired with a 20 cm focal length OAP1 for large-area imaging on the millimeter scale (Fig. 2a-d, 3a-c), resulting in a wide mid-IR illumination profile. The optical layout was changed for low-wavenumber measurements below 900 $cm^{-1}$ to achieve optimal beam diameter at the sample plane (Fig. 2e-f). In this configuration, OAP1 was replaced with a 5 cm focal length OAP1 to achieve tighter focusing at longer wavelengths. Furthermore, OAP3 and mirror M5 were flipped down to redirect the beam path towards M6. An OAP1 with a 15 cm focal length was used to image tuberculosis-infected tissue sections (Fig. 3d-f). This intermediate focal length provides a balance between FOV and illumination intensity, enabling sufficient photothermal excitation across the tissue while minimizing thermal damage. The mid-IR laser power is adjustable via the laser's built-in attenuation controls.

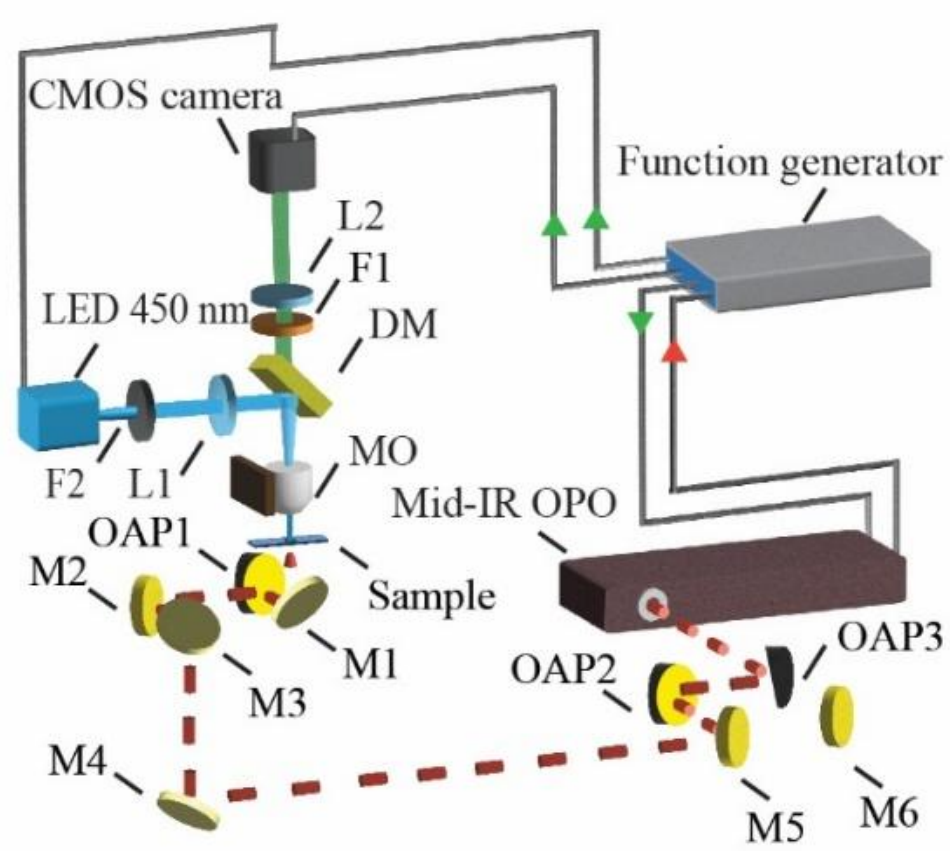

Fig. 1. Optical and electronic layout of the wide-field F-MIP microscope. LED illumination is filtered (F2) and relayed by lens L1 to the back focal plane of the microscope objective (MO). The excitation light is reflected by a dichroic mirror (DM) onto the sample. Fluorescence emission is collected by the MO, transmitted through the DM and emission filter (F1), and detected by a CMOS camera. The mid-IR beam is directed to the sample plane by a series of mirrors and off-axis parabolic mirrors. A function generator synchronizes the mid-IR pulse train with the visible LED and the CMOS camera triggers to acquire alternating “hot (mid-IR-on)” and “cold (mid-IR-off)” frames.

The LED (1 µs pulse duration), mid-IR laser, and camera are operated in a pulsed and synchronized manner. The internal trigger of the mid-IR OPO is synchronized with a function generator (9214+, Quantum composers). Using this internal trigger as a reference, the function generator provides external trigger signals to the visible probe source, the camera, and a gate signal to the mid-IR laser. The gate signal controls the on/off modulation of the OPO output. The resulting on/off (hot/cold) images are acquired and processed using NI LabVIEW 2025.

Fluoromax Green PS particles (5 µm in diameter), purchased from Thermo Fisher Scientific, were first prepared as a stock suspension, then diluted into 25 mL of deionized water. A 6 µL aliquot of this diluted suspension was deposited onto a $CaF_2$ or a float-zone Si substrate (200 µm thick) by pipetting and allowed to air-dry before imaging. Preparation of tuberculosis-infected tissue was explained in our previous publication[13].

Wide-field F-MIP images of fluorescent PS beads were first acquired to assess the imaging performance of the system. Imaging was performed with a 10× objective at 5 fps, and 25 consecutive frames were averaged to improve the SNR. The resulting image recorded at 3026 $cm^{-1}$ (Fig. 2a) shows an almost circular FOV of ~1 mm diameter. A comparison of the achievable FOV in the QCL-based F-MIP setup is shown in Fig. S1, where a wide-field F-MIP image of PS beads at 1450 $cm^{-1}$ yields a FOV of approximately 55 µm. The vibrational response at 3026 $cm^{-1}$ corresponds to the aromatic C–H stretching mode of the PS phenyl rings, which produces strong absorption and therefore high photothermal contrast. A smaller region of interest, highlighted by a green rectangle in Fig. 2a, was selected to investigate spectral contrasts at different wavenumbers (Fig. 2b). At 2850 $cm^{-1}$ (Fig. 2c), the signal primarily arises from aliphatic C–H symmetric stretching vibrations, whose absorbance is weaker in PS than at 3026 $cm^{-1}$, leading to reduced image contrast. At 2980 $cm^{-1}$ (Fig. 2d), imaging occurs in an off-resonance state with minimal vibrational absorption, resulting in negligible photothermal contrast. Multiple F-MIP images acquired at 10 $cm^{-1}$ intervals across the 2800–3100 $cm^{-1}$ range were used to construct a spectrum (Fig. 2g). For reference, the FTIR spectrum of PS is provided in Supplementary Fig. S2. The F-MIP spectrum demonstrates good spectral fidelity in comparison with the FTIR reference.

The imaging performance of our F-MIP setup surpasses the FOV and wavelength coverage reported in earlier wide-field implementations. As summarized in Table 1, prior studies typically achieved FOVs of 700 µm×400 µm (an area of $2.2\times10^5$ $\mu m^2$) using sensitive phase-shifting interferometric imaging[14]. In contrast, our system reached 1 mm diameter ($7.9\times10^5$ $\mu m^2$) when imaging PS microspheres in the high-wavenumber region (Fig. 2). This approximately 3.6 times larger area compared with the state of the art and more than 280 times larger than the QCL-based F-MIP (see Fig. S1) which arises mainly from the high pulse energy of the newly developed single-housing OPO laser, which exceeds 360 µJ per pulse at the source (~280 µJ at the sample plane) in the 2500–3200 $cm^{-1}$ spectral range (Fig. S3). No previously reported wide-field MIP source – QCL, FEL, or OPO – has exceeded 80 µJ per pulse energy[13-15]. Wide-field MIP systems using QCLs and free-electron lasers (FELs) as pump sources often operate at high repetition rates[13, 16, 17], which could limit the thermal relaxation time compared to a high-energy OPO source. QCL systems typically delivering pulse energies around 100 nJ or less, achieve smaller FOVs than FEL systems providing higher pulse energies up to 30 µJ[13, 18]. Moreover, earlier OPO-based systems offered high pulse energies (up to 80 µJ) but with narrower tuning ranges[14].

The millimeter-scale FOV was achieved using a 0.25 NA objective, yielding a spatial resolution of 1.36 µm (Fig. S4), consistent with the diffraction limit at 500 nm fluorescence emission (0.61λ/NA); as the resolution depends on the NA of the objective, it can be further improved by employing higher-NA objectives, as demonstrated in Fig. S4. The wide-field chemical imaging capability of our F-MIP setup could be particularly suitable for applications such as rapid detection and classification of micro- and nanoplastics in environmental and biological samples[19, 20].

To assess the thermal impact of OPO excitation, we calibrated the temperature sensitivity of Fluoromax Green fluorescence (−1.31%/°C, Fig. S5) and used it to convert the fluorescence change observed during photothermal imaging into a local temperature rise. A pulse energy of 280 µJ delivered over an area of $7.9\times10^5$ $\mu m^2$ corresponds to a fluence of 354 pJ/$\mu m^2$. At 3026 $cm^{-1}$, this yielded an estimated temperature rise of 4.6 K (Fig. S5) well below the glass transition temperature of PS enabling nondestructive imaging. Furthermore, a previous study demonstrated nondestructive imaging of single cells using a fluence of 1.1 nJ/$\mu m^2$ [10].

The imaging performance was further assessed in the fingerprint region at 698 $cm^{-1}$ and 756 $cm^{-1}$, in addition to the C–H stretching region. Measurements were conducted using a 50×/0.75 NA objective lens, resulting in a FOV of approximately 240 µm×220 µm. As $CaF_2$ is not transparent at these wavenumbers anymore, the PS beads were prepared on float-zone silicon (Si) substrates (transmission of 55-60% at 13.2 µm and 14.2 µm, as measured with a S470C (Thorlabs) sensor), chosen for their high optical purity and minimal absorption at longer mid-IR wavelengths. The image (Fig. 2e) obtained at 698 $cm^{-1}$ corresponds to the aromatic C–H out-of-plane bending mode of the monosubstituted benzene ring in PS. Analogously, the signal observed at 756 $cm^{-1}$ (Fig. 2f) originates from another characteristic out-of-plane C–H bending vibration of the phenyl ring.

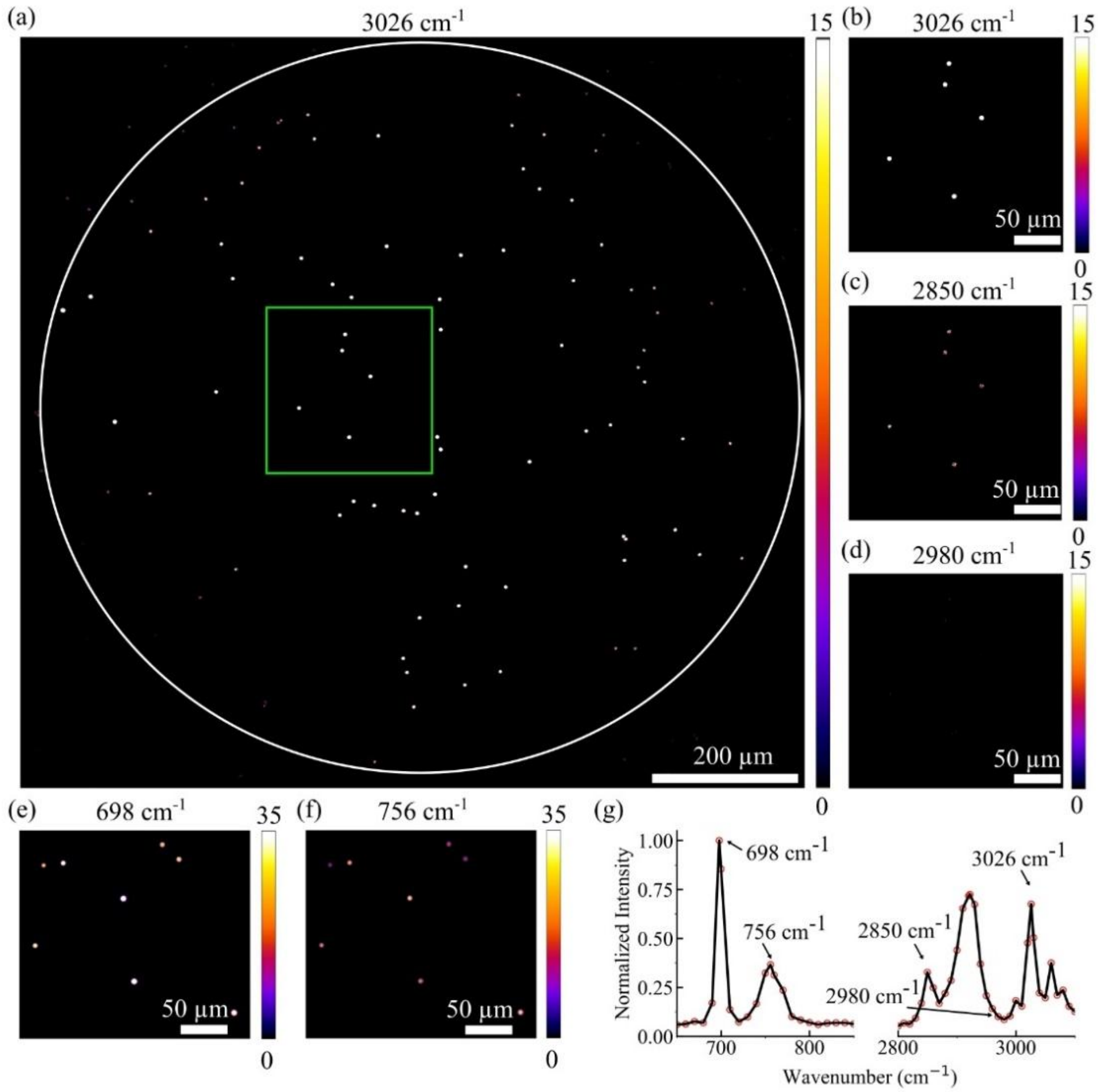

**Fig. 2.** F-MIP multispectral imaging of 5 µm PS beads. (a) Wide-field F-MIP image at 3026 cm$^{-1}$, with the FOV dimensions of ~1 mm diameter (white circle). (b) Magnified view of the green rectangular region marked in (a). (c-d) Wide-field F-MIP contrast at 2850 cm$^{-1}$ and 2980 cm$^{-1}$ extracted from the same green rectangular region. Pulse energy of OPO at the sample plane was ~280 µJ (2500 – 3200 cm$^{-1}$ range). (e) F-MIP image at 698 cm$^{-1}$. (f) F-MIP image at 756 cm$^{-1}$. Pulse energy of OPO at the sample plane was ~20 µJ (at 698 and 756 cm$^{-1}$). (g) Spectra acquired from hyperspectral imaging of F-MIP images across two wavenumber ranges, 650–850 cm$^{-1}$ and 2800–3100 cm$^{-1}$, with a wavenumber interval of 10 cm$^{-1}$.

While prior wide-field MIP (see Table 1) was mostly limited to wavenumbers above 950 cm$^{-1}$ [15], our configuration extended the range deep into the fingerprint region (Fig. 2e-f), allowing F-MIP imaging at wavenumber 698 cm$^{-1}$ (down to 625 cm$^{-1}$). Even in this regime, where the pulse energy decreases to about 40 µJ (~20 µJ at the sample plane), the available pulse energy remains higher than for most mid-IR sources, supporting imaging FOVs exceeding 200 µm. Imaging in this spectral range, however, requires an IR-transparent substrate such as float-zone silicon, since standard substrates like $CaF_2$ become absorptive below ~1000 cm$^{-1}$. Pump–probe delay imaging of PS beads on $CaF_2$ and Si substrates showed this heating relaxes back to equilibrium with decay constants of ~16.8 and 16.4 µs, respectively, fully dissipating within ~80–100 µs (Fig. S6). Similar decay constants across both substrates likely reflect the small contact area between the spherical beads and the flat substrate, which constrains heat dissipation regardless of the thermal conductivity of the substrate material. This thermal relaxation timescale corresponds to a pulsed laser repetition rate of ~11.1 kHz ($1/90\times10^{-6}$ s). For particles larger than 5 µm, the decay time would increase due to greater thermal mass[10]. Wide-field MIP heterodyne imaging of potassium ferricyanide microparticles showed a thermal relaxation time on the order of milliseconds[8], implying the requirement of a repetition rate below 1 kHz to fully resolve the thermal decay dynamics. In our OPO source, the 100 Hz repetition rate yields a much longer pulse interval (>10 ms), providing ample time for even larger samples to cool between mid-IR exposures. This combination of high pulse energy, low repetition rate, and broad spectral tuning in a single, commercially available laser source has not, to our knowledge, been demonstrated previously.

**Table. 1.** Summary of previously reported wide-field MIP imaging systems, highlighting their mid-IR sources, photothermal contrast mechanisms, and achieved FOV.

| Laser source | Contrast mechanism | FOV | Pulse energy and width or power | Repetition rate | Tuning range | Resolution |
|---|---|---|---|---|---|---|
| OPO[6] | Scattering field | ~35 µm | 150 nJ, <50 ns | 20 kHz | 1176 – 1785 $cm^{-1}$ | <15 $cm^{-1}$ |
| QCL[7] | Optical phase | 100 µm × 100 µm | ~100 nJ, 1 µs | 1 kHz | 1450 -1640 $cm^{-1}$ | <1 $cm^{-1}$ |
| QCL[21] | Optical phase | 100 µm × 100 µm | 60 mW | Continuous | 1000-1111 $cm^{-1}$ | 2.5 $cm^{-1}$ |
| QCL[15] | Optical phase | 460 µm × 460 µm | 500 mW[a] | 600 Hz | 950 – 1800 $cm^{-1}$ | <1 $cm^{-1*}$ |
| QCL[16] | Dark-field scattering | 70 µm× 70 µm | ~12 mW, ~1 µs | 200 kHz | 1500 – 1700 $cm^{-1}$ | <1 $cm^{-1}$ |
| EC-QCL[8] | Scattering field | 120 µm | 160 mW peak | 100 kHz | 2015 -2220 $cm^{-1}$ | <1 $cm^{-1*}$ |
| OPO[9] | Optical phase | 186 µm | 80 µJ, 9 ns | 1 kHz | 2500 - 4475 $cm^{-1}$ | <10 $cm^{-1}$ |
| OPO[22] | Fluorescence | 60 µm × 60 µm | 20 mW, 50 ns | 20 kHz | 1175 – 1800 $cm^{-1}$ | <15 $cm^{-1}$ |
| OPO[10] | Optical phase | 69 µm × 69 µm | 10 µJ, 6 ns | 1 kHz | 2800 – 3250 $cm^{-1}$ | ~10 $cm^{-1}$ |
| OPO[23] | Optical phase | ~10 µm | 1.7 µJ, 0.5 ns | 1 kHz | 2650 – 3150 $cm^{-1}$ | ~10 $cm^{-1}$ |
| OPO[14] | Optical phase | 700 µm × 400 µm | 80 µJ, 9 ns | 1 kHz | 2500 - 4475 $cm^{-1}$ | <10 $cm^{-1}$ |
| FEL[13] | Scattering field | 240 µm × 170 µm | 1 µJ, <30 ps | 13 MHz | 1300 – 1800 $cm^{-1}$ | <15 $cm^{-1}$ |
| **OPO (our study)** | Fluorescence | 1030 µm × 1010 µm | Up to 360 µJ, ~26 ps | 100 Hz | 625 – 4347 $cm^{-1}$ | <4 $cm^{-1}$ |

Values marked with [a] are the author's approximations derived from source data (not explicitly stated).

Wide-field F-MIP imaging was next applied to tuberculosis-infected tissue sections, using intrinsic tissue fluorescence[24] as the reporter signal. As tuberculosis pathology is spatially heterogeneous, large contiguous FOVs are needed to capture granulomas and lipid-laden macrophages without small-FOV sampling bias. F-MIP images of a tuberculosis-infected tissue section were acquired at 2920 $cm^{-1}$ using 10×/0.25 NA objective, corresponding to the C–H stretching vibration of lipids, over a 1050×990 µm FOV (Fig. 3b). To capture larger tissue areas, we tiled nine such fields (756×756 µm each, accounting for overlap) into a 3×3 mosaic spanning more than 2.2 mm, with each frame acquired at 1 fps and averaged over five photothermal images. The F-MIP image was normalized (hot-cold/cold) to correct local variations in fluorescence. Individual tiles were stitched using a previously described algorithm[25], with a 3% overlap region between adjacent tiles used for feature matching. The intensity decrease toward the tile edges arises from the Gaussian beam profile of the mid-IR excitation, as confirmed in Fig. S7c. Imaging the same sample with a 20×/0.4 NA objective showed that the F-MIP image (Fig. S7d) fills the entire camera sensor (~14.6×12.6 mm, 2664×2304 pixels at an effective 5.48 µm pixel size after 2×2 binning) while retaining a spatial resolution below 1 µm (Fig. S4).

Furthermore, a 7×5 mosaic was acquired in the fingerprint range (tile size 162 µm×227 µm; Fig. S7e); to minimize contrast loss at tile boundaries from the Gaussian IR beam profile (Fig. S7f), a smaller tile size (130 µm×162 µm) was used for the main dataset (stitched images, Fig. 3d). Vibrational contrast at 1660 $cm^{-1}$ and 1740 $cm^{-1}$ resolves protein and lipid distributions, respectively: the amide I band (C=O stretching of peptide bonds) maps overall protein content (Fig. 3d), while the carbonyl stretching signature at 1740 $cm^{-1}$ is consistent with lipid accumulation associated with tuberculosis pathology (Fig. 3e)[26, 27]. Overlaying these two channels (cyan, 1660 $cm^{-1}$; red, 1740 $cm^{-1}$; Fig. 3f) reveals spatially distinct lipid-enriched regions within the protein-rich tissue background, consistent with the presence of mycobacteria or lipid-laden macrophages mounting an immune response to infection.

Beyond technical performance, the biological imaging results demonstrate the practical impact of this approach. MIP imaging has previously been used to study biological tissue sections[15, 28]. Imaging of tissue sections at 1660 and 1740 $cm^{-1}$ revealed characteristic spectral contrast between regions of interest and surrounding tissue, although the 1740 $cm^{-1}$ signal may reflect contributions from both infection-associated components and endogenous lipids [26, 27, 29]. Resolving this lipid–protein relationship over large, contiguous areas, rather than isolated fields, is key to characterizing the extent of infection, and is achieved here with substantially less tiling and acquisition time than would be required with previously reported wide-field MIP FOVs (Table 1).

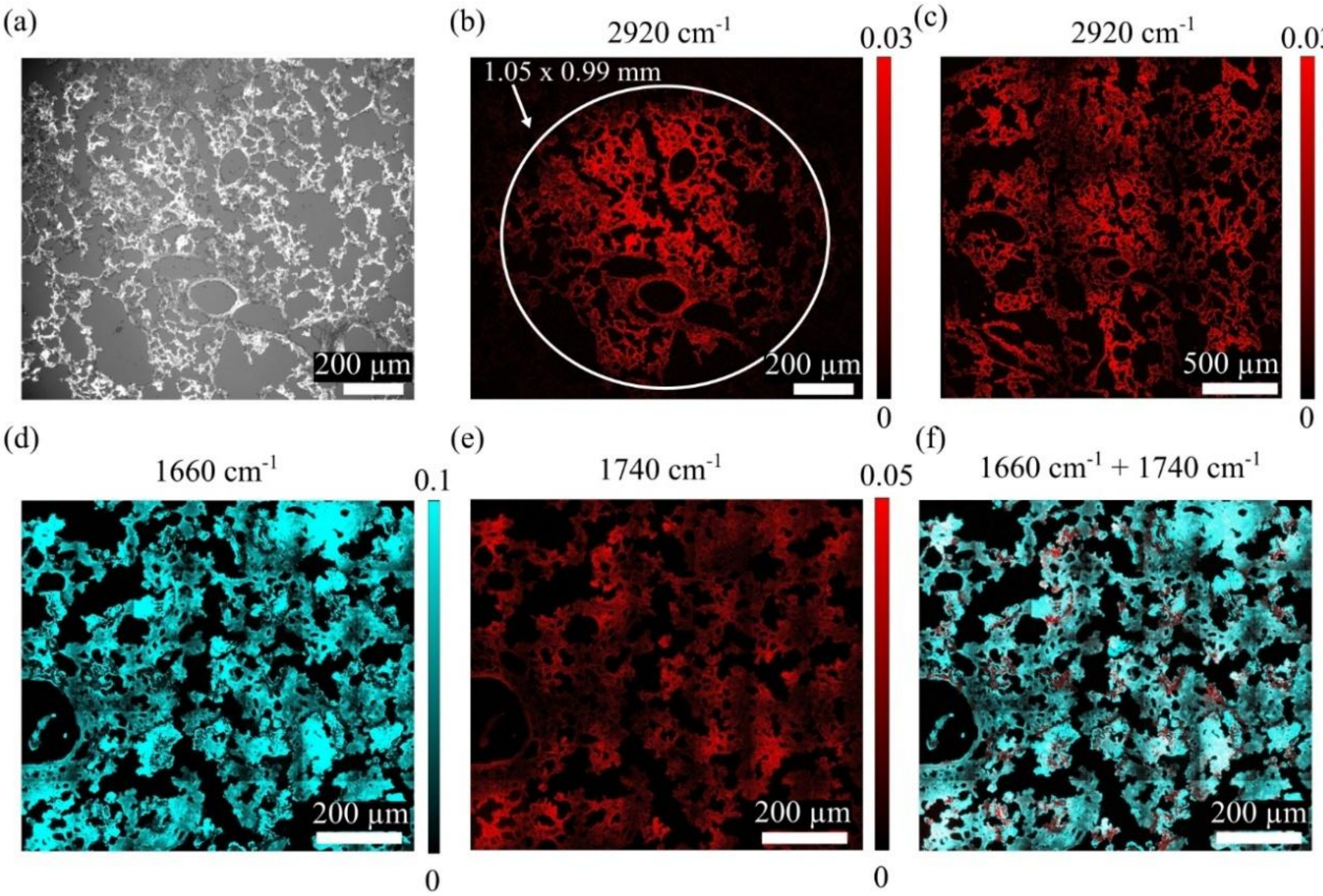

**Fig. 3.** Fluorescence and wide-field F-MIP imaging of a tuberculosis-infected tissue section. (a) Fluorescence image. (b) F-MIP image at 2920 cm$^{-1}$. (c) 3×3 mosaic F-MIP image at 2920 cm$^{-1}$, spanning an area exceeding 2 mm. Pulse energy at the sample plane was ~280 µJ for (b) and (c). (d,e) Wide-field F-MIP images of a different region of the same tissue section at 1660 cm$^{-1}$ and 1740 cm$^{-1}$, respectively. (f) Overlay of the images in (d) and (e). Pulse energy at the sample plane was <30 µJ for (d–f).

A FOV exceeding 220 µm (major axis of the ellipse in Fig. S7) was achieved for these measurements at 1660 and 1740 cm$^{-1}$, even though only a fraction of the available laser energy (about 30 µJ out of more than 130 µJ) was used (Fig. S3). These results underscore the system's efficiency and suggest substantial capacity to scale the illuminated area through further beam optimization. The overlay of the 1660 cm$^{-1}$ and 1740 cm$^{f}$ absorption maps highlighted regions of spectral contrast; future studies will employ additional biomarkers to distinguish lipid signals from infection-specific features, thereby isolating true bacterial signatures. The demonstrated capability of wide-field F-MIP to acquire large-area, chemically resolved images from biological tissue sections underscores its potential for label-free biomolecular mapping, composition analysis, and diagnostic applications.

Despite these advantages, a few limitations should be recognized. Fluorescence-detected F-MIP relies on endogenous or exogenous fluorophores as the reporter signal, which may limit applicability to samples with insufficient intrinsic fluorescence or that cannot be labeled. Additionally, the relatively low repetition rate of the OPO, while beneficial for achieving high pulse energy and allowing sufficient thermal relaxation, constrains the achievable frame rate of the CMOS camera and thus the imaging speed. The illumination across the F-MIP FOV is not yet fully homogeneous: the Gaussian beam profile of the pump source reduces signal uniformity toward the edges of the images (Fig. 3), slightly lowering photothermal contrast across the extended field. This can affect the quantitative evaluation of large mosaics or heterogeneous areas. Future optimization of the focal conditions by adjusting the OAP1 focal length (Fig. 1) or homogenizing the beam (top-hat beam profile) can flatten intensity profiles and further increase the usable FOV. Additionally, the OPO shows a reduction in pulse energy from shorter to longer wavelengths, decreasing from about 360 µJ to 15 µJ (Fig. S3) across the tuning range (4347–625 cm$^{-1}$). Implementing mid-IR pulse energy stabilization could help maintain consistent output during wavelength tuning, which would be particularly beneficial for improving sensitivity and data quality in multispectral measurements. Another important direction involves improving the consistency of focusing conditions across the broad mid-IR tuning range. In the current setup, the strong wavelength dependence of the beam divergence requires OAP1 to be manually repositioned when switching between widely spaced wavenumbers. Developing an adjustable or automated beam-conditioning module that maintains a constant illumination size at the sample plane would eliminate this requirement, ensure repeatable alignment, and facilitate multispectral imaging across a broad spectral range. Addressing these technical aspects will further enhance performance and broaden the applicability of F-MIP beyond the tuberculosis and polystyrene demonstrations shown here, potentially extending to infectious disease diagnostics, digital histopathology, and materials science, including microplastic characterization.

### Supplementary document

See Supplement 1 for supporting content.

### Author information

### Corresponding author

**Christoph Krafft** - *Leibniz Institute of Photonic Technology, Member of Leibniz Research Alliance Leibniz Health Technologies, Member of the Leibniz Center for Photonics in Infection Research, Albert-Einstein-Str. 9, 07745 Jena, Germany.*

### Authors

**Anooj Thayyil Raveendran** - *Leibniz Institute of Photonic Technology, Member of Leibniz Research Alliance Leibniz Health Technologies, Member of the Leibniz Center for Photonics in Infection Research, Albert-Einstein-Str. 9, 07745 Jena, Germany.*

**Supatcharee Cael** - *Leibniz Institute of Photonic Technology, Member of Leibniz Research Alliance Leibniz Health Technologies, Member of the Leibniz Center for Photonics in Infection Research, Albert-Einstein-Str. 9, 07745 Jena, Germany.*

**Anju Jose -** *Otto Schott Institute of Materials Research, Friedrich-Schiller-University Jena, Lessingstrasse 14, 07743 Jena, Germany*

**Jürgen Popp** - *Leibniz Institute of Photonic Technology, Member of Leibniz Research Alliance Leibniz Health Technologies, Member of the Leibniz Center for Photonics in Infection Research, Albert-Einstein-Str. 9, 07745 Jena, Germany. Friedrich-Schiller-University Jena, Institute of Physical Chemistry, Member of Leibniz Research Alliance Leibniz Health Technologies, Member of the Leibniz Center for Photonics in Infection Research, Helmholtzweg 4, 07743 Jena, Germany.*

### Author contribution

A.T.R. contributed to the conception of the study, developed the F-MIP setup, and performed the measurements and software work. S.C. assisted with measurements and provided feedback on the manuscript. A. J assisted with the measurement of fluorescent dependence on temperature. C.K. supervised the project, reviewed the manuscript, acquired funding, and contributed to revisions. J.P. supervised the project, acquired funding, and reviewed the manuscript. A.T.R. drafted the manuscript with input from all authors. All authors contributed to manuscript revision and approved the final version.

### Disclosures

The authors declare no conflicts of interest.

### Data availability

Data supporting the results in this paper is not publicly available at this time but is available from the corresponding author upon reasonable request.

### ACKNOWLEDGMENT

This work was supported by the Federal Ministry of Research, Technology, and Space (BMFTR) through the funding program *Leibniz Center for Photonics in Infection Research (LPI, FKZ 13N15464)*. The authors gratefully acknowledge N. Redinger (FZ Borstel, Germany) for the preparation of the infected murine lung tissue samples used in this study. We thank P. Susanne, R. Cornelia, and Samir F. El-Mashtoly for providing the *E. coli* bacteria used in the spatial resolution measurements, and L. Wondraczek for generously providing access to their instrument for the fluorescence temperature dependence measurements.